\documentclass[prx,superscriptaddress,amsfonts,amssymb,amsmath,floats,twocolumn,aps,footinbib,shownopacs]{revtex4}
\usepackage[pdftex]{graphicx}
\usepackage{subfigure}
\usepackage{dsfont}
\usepackage{lipsum}
\usepackage{dcolumn}
\usepackage{bm}
\usepackage{color}
\usepackage{physics}
\usepackage{multirow}
\usepackage[pdftex,colorlinks=true, linkcolor=blue,citecolor=blue,filecolor=blue]{hyperref}
\usepackage{systeme}

\usepackage{comment}
\usepackage{bbold}

\usepackage{soul}

\usepackage{array}
\newcommand{\PreserveBackslash}[1]{\let\temp=\\#1\let\\=\temp}
\newcolumntype{C}[1]{>{\PreserveBackslash\centering}p{#1}}
\newcolumntype{R}[1]{>{\PreserveBackslash\raggedleft}p{#1}}
\newcolumntype{L}[1]{>{\PreserveBackslash\raggedright}p{#1}}
\usepackage{pifont}

\begin{document}
\title{Nonequilibrium control of kagome metals}

\author{Francesco Grandi} 
\affiliation{Institut f\"ur Theorie der Statistischen Physik, RWTH Aachen University, 52056 Aachen, Germany}

\author{Ronny Thomale}
\affiliation{Institut f\"ur Theoretische Physik und Astrophysik and W\"urzburg-Dresden Cluster of Excellence ct.qmat, Universit\"at W\"urzburg, 97074 W\"urzburg, Germany }\affiliation{ Department of Physics and Quantum Centers in Diamond and Emerging Materials (QuCenDiEM) group, Indian Institute of Technology Madras, Chennai 600036, India}

\author{Dante M. Kennes}
\affiliation{Institut f\"ur Theorie der Statistischen Physik, RWTH Aachen University, 52056 Aachen, Germany}
\affiliation{JARA-Fundamentals of Future Information Technology, 52056 Aachen, Germany}
\affiliation{Max Planck Institute for the Structure and Dynamics of Matter, Center for Free-Electron Laser Science (CFEL), Luruper Chaussee 149, 22761 Hamburg, Germany}

\begin{abstract}
Exotic quantum order in kagome metals, i.e., quantum materials with a Fermi liquid parent state of electrons on a kagome lattice, has appeared as a vibrant emerging field of condensed matter physics. Already in a small kagome material subclass such as vanadium-based compounds $A$V$_3$Sb$_5$ ($A=$ K, Rb, Cs), the first wave of experimental exploration has brought about manifold evidence for hitherto largely elusive phenomena such as high-temperature charge ordering with orbital currents, nematic order, cascades of charge ordering transitions with hierarchies of ordering vectors, and unconventional superconductivity. We argue that kagome metals promise to be a prototypical ground for the non-equilibrium analysis of quantum order through time-dependent parameter control and manipulation. In particular, we propose to investigate the nematic character of kagome quantum order through light and strain pulses, as well as the nature of time-reversal symmetry breaking and chirality through properly polarized laser pulses.
\end{abstract} 
\maketitle

\section{Introduction} \label{sec:intro}
Recent advances in nonequilibrium control of matter, and especially of light-control of materials' properties, raised high expectations about future technological applications in the field of ultrafast quantum materials science \cite{Forst2011_NatPhys,Giannetti2016_AdvPhys,Basov2017_NatMat,delaTorre2021_RMP,Marino2022_RepProgPhys,Murakami2023_arXiv}. Besides the potential technological impact, many experiments pose tantalizing and challenging theoretical questions, e.g. concerning the meaning of the free energy landscape usually employed to describe the dynamics of the order parameter at a photoinduced melting and the subsequent recovery of the symmetry broken state \cite{Wall2018_Science,Perez-Salinas2022_NatComm,delaPenaMunoz2023_NatPhys,Picano2023_PRB} or the role of dissipation and multi-cycle pulses to dress electronic states \textit{\`a la} Floquet engineering \cite{Ito2023_Nat}. Finding suitable material platforms for the realization of these time-dependent studies is of prime relevance both because they might help answering the above raised questions and because they might yield important progress on the subject of control or even the induction of exotic and unconventional states of matter. In this regard, the recently discovered kagome metals are particularly promising.

\begin{figure}[h!]
    \centerline{\includegraphics[width=0.49\textwidth]{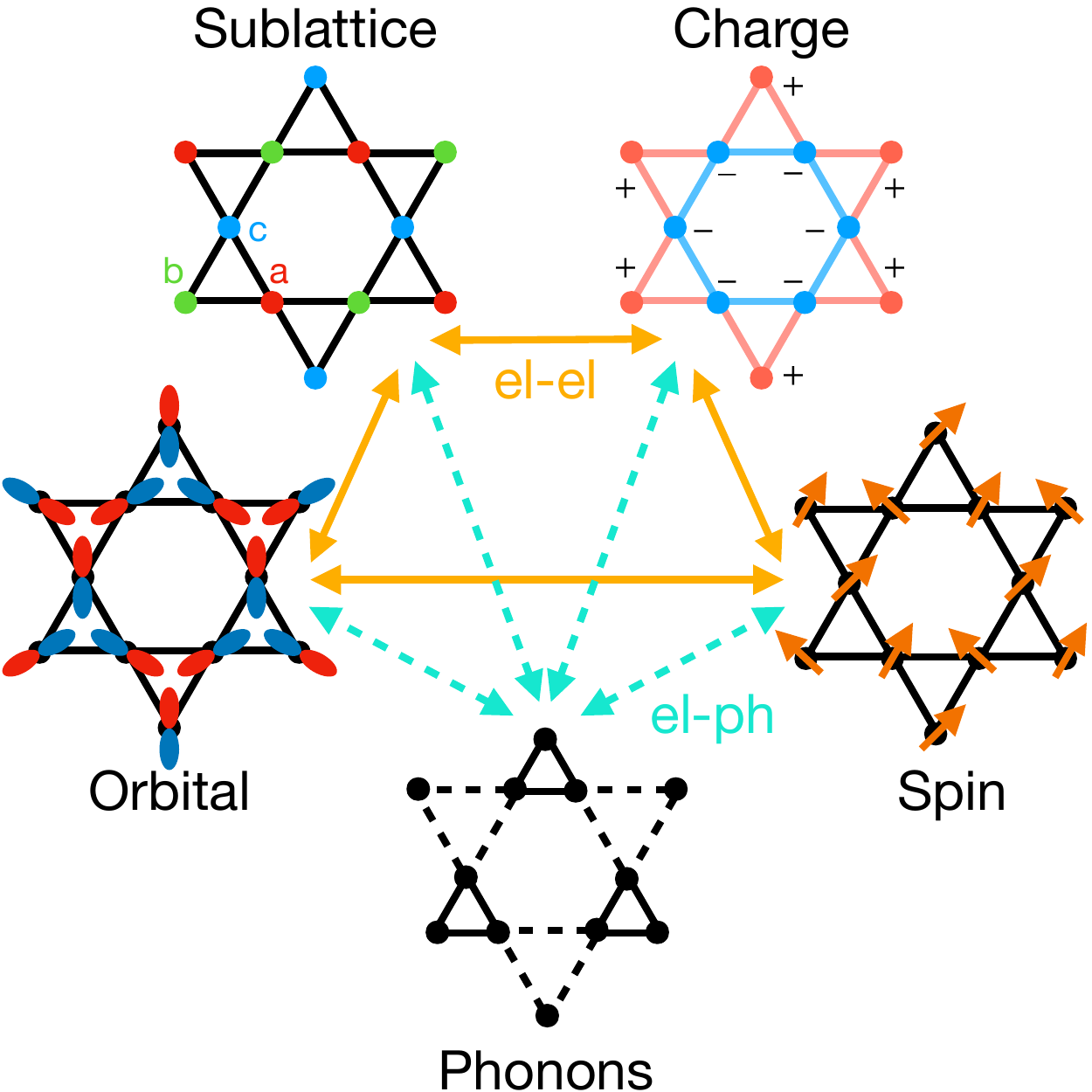}}
	\caption{\textbf{Intertwined degrees of freedom and corresponding interactions in the kagome metals -} The sublattice, the charge, the orbital, the spin and the phononic degrees of freedom conspire to give rise to the unusual physics of the kagome metals. The continuous yellow (dashed celeste) double arrows connecting the different degrees of freedom represent the electron-electron (electron-phonon) interaction.
    } \label{fig:interactions}
\end{figure}

\noindent
The unusual states observed in the kagome metals include cascades of charge ordering transitions \cite{Zhao2021_Nat,Zheng2022_Nat} with broken time-reversal symmetry \cite{Mielke2022_Nat} and chiral signatures \cite{Jiang2021_NatMat,Guo2022_Nat}, nontrivial topology \cite{Yang2020_SciAdv}, nematicity \cite{Li2022_Nat_Phys,Nie2022_Nat,Asaba2024_NatPhys} and unconventional superconductivity \cite{Chen2021_Nat}. The stabilization of these exotic phases is believed to arise from a subtle interplay between the charge, the orbital, the spin, the sublattice and the phononic degrees of freedom, each of them talking with each other thanks to the electron-electron or the electron-phonon interactions, both of which should be relevant in this experimental platform as schematically depicted in Fig.~\ref{fig:interactions}. The equilibrium phase diagram of the kagome metals is still strongly debated, with contradictory experimental reports regarding the presence of the above-mentioned symmetry broken states, suggesting a highly nontrivial energy landscape for these compounds with metastable minima almost degenerate with the ground state, the balance of which can be detuned with small perturbations \cite{Guo2024_NatPhys}. For this reason, kagome metals present a perfect playground for nonequilibrium physics since one might easily induce metastable states by external driving, and it might be possible to learn more about the equilibrium properties of these materials using out-of-equilibrium routes.

\begin{figure}
    \centerline{\includegraphics[width=0.49\textwidth]{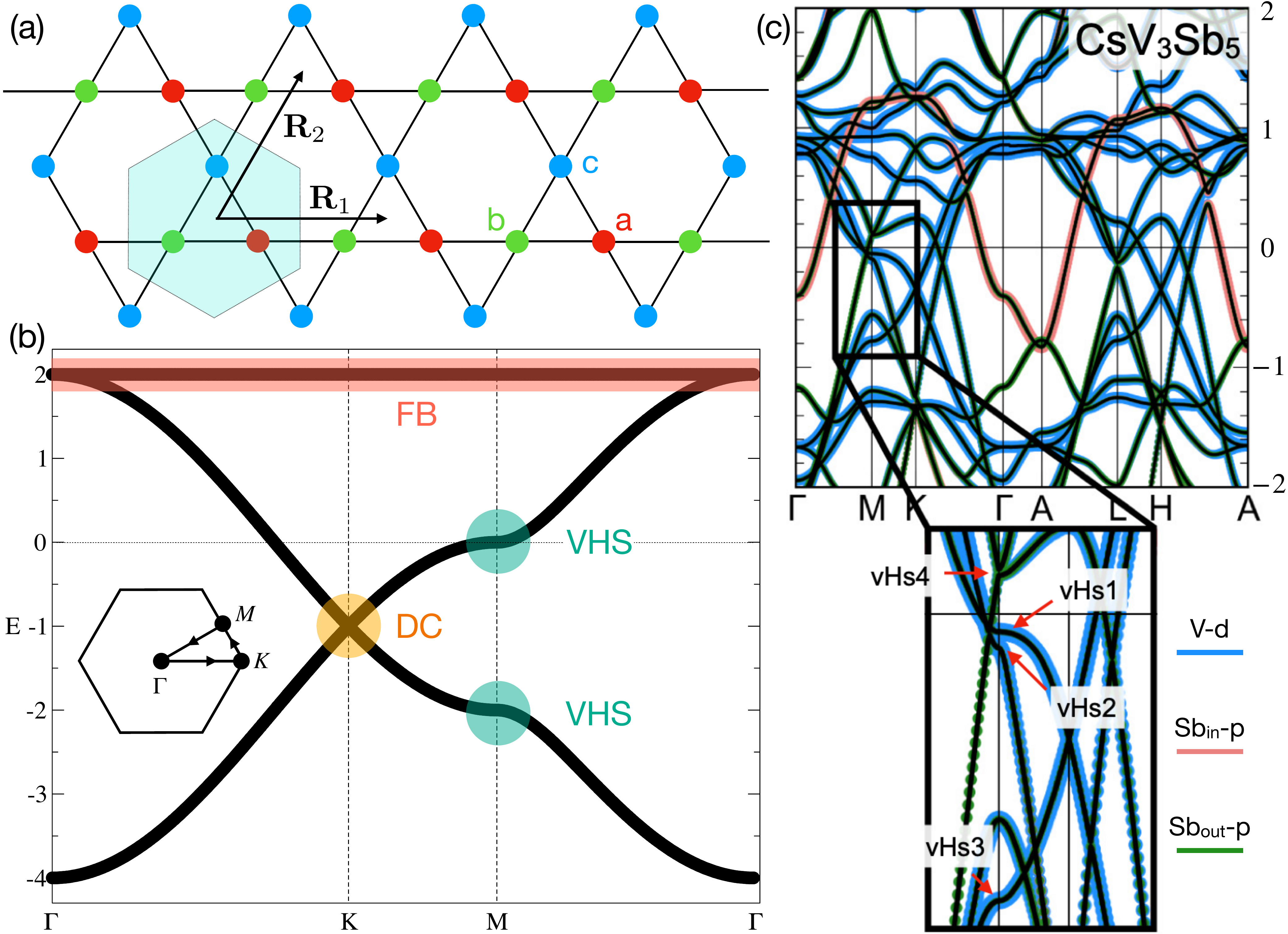}}
	\caption{\textbf{Kagome lattice, tight-binding and \textit{ab initio} dispersion relations of the kagome metals -} (a) The kagome lattice with its hexagonal Wigner-Seitz cell and the three sublattices a,b and c. The vectors $\mathbf{R}_1$ and $\mathbf{R}_2$ are the unit vectors. (b) Band dispersion of the kagome lattice as obtained from a tight-binding Hamiltonian with nearest-neighbor hopping (hopping strength $t=1$). The inset shows the first Brillouin zone and the path along which the energy dispersion is shown. The Dirac cone (DC), the upper and the lower van Hove singularities (VHSs) and the flat band (FB) are highlighted in color. (c) Dispersion relation (vertical axis in eV) obtained from \textit{ab initio} simulations for CsV$_3$Sb$_5$, adapted from \cite{LaBollita2021_PRB}. The color blue indicates that the corresponding band has mainly a contribution coming from the d vanadium orbitals, while the orange (green) bands have mainly a contribution coming from the in-(out-)of-plane antimony atoms. Several of the features marked in panel (b) can be recognized in panel (c). The lower panel shows a zoom of the region near the M and the K points, where several VHSs and DCs are present near the Fermi level.}
    \label{fig:kag_lat_disp}
\end{figure}

\noindent
In particular, we focus on the class of the vanadium-based kagome metals \cite{Ortiz2019_PRM}, which hosts the three family members $A$V$_3$Sb$_5$ ($A =$K, Rb, Cs) \cite{Neupert2022_NatPhys,Jiang2022_NSR,Wilson2024_NatRevMat}. The kagome metals find in the kagome lattice displayed in Fig.~\ref{fig:kag_lat_disp}(a) their basic building block since these three-dimensional materials are made of parallel kagome layers formed by the vanadium atoms. Already the nearest neighbor hopping tight-binding electronic band structure is extremely intriguing, hosting two Dirac cones (DCs) away from the Fermi level at the points K and K' in the Brillouin zone, two van-Hove singularities (VHSs) at each of the three independent M points and a flat band (FB) over the entire Brillouin zone \cite{Kiesel2012_PRB}, see Fig.~\ref{fig:kag_lat_disp}(b).

\noindent
Despite the fact that the band structure of the kagome metals is actually much more complicated than the one of the single orbital kagome lattice, still one can recognize several of the main features identified above in the \textit{ab initio} electronic band structure. In the pristine high-temperature phase, polarization-dependent angle-resolved photoemission spectroscopy (ARPES) \cite{Hu2022_NatComm} and density functional theory calculations \cite{LaBollita2021_PRB,Consiglio2022_PRB} find that most of the states near the Fermi level come from the d orbitals of the vanadium atoms, which form VHSs, DCs and almost FB at low-energies, see Fig.~\ref{fig:kag_lat_disp}(c).

\noindent
At T$_\text{co} \sim 90$K, an unconventional charge order (CO) with in-plane reconstruction has been measured for the three compounds together with the onset of an anomalous Hall effect \cite{Yang2020_SciAdv,Yu2021_PRB,Wang2023_JouPhysMat}. The CO might have an out-of-plane component, but the precise form of it remains controversial \cite{Li2022_NatComm,Jiang2021_NatMat,Li2021_PRX,Liang2021_PRX,Ortiz2021_PRX}. Much of the interest concerning the CO phase comes from the experimental reports of time-reversal symmetry (TRS) breaking without any signatures of local magnetic moments  \cite{Yang2020_SciAdv,Yu2021_PRB,Neupert2022_NatPhys,Khasanov2022_PRR,Mielke2022_Nat}, which might be explained by the presence of orbital currents \cite{Nayak2000_PRB}, in a way similar to what is found theoretically  for the honeycomb (Lieb) lattice in the Haldane \cite{Haldane1988_PRL} (Varma \cite{Varma1997_PRB}) model. The actual presence of TRS breaking remains controversial \cite{Saykin2023_PRL}, with contradictory experimental findings which observed a broken TRS even above the onset of the CO \cite{Asaba2024_NatPhys}, in a region where just the fluctuations of the lower temperature CO are observed \cite{Chen2022_PRL,Subires2023_NatComm,Yang2023_PRB}.

\noindent
At T$_\text{nem} \sim 30-50$K, several experiments report the transition to a nematic state where each kagome layer might have a lower rotational symmetry than the one of the lattice \cite{Nie2022_Nat,Li2022_NatComm}. In a comparable temperature range, the onset of a one-dimensional CO that simultaneously breaks the rotational and the discrete translational symmetries of the higher energy CO has been observed in some experiments \cite{Zhao2021_Nat,Guo2022_Nat,Li2022_NatComm,Li2022_PRB,Li2023_PRX}. In contrast, other experiments do not find any transition at T$_\text{nem}$ \cite{Liu2024_PRX,Frachet2023_arXiv}, suggesting that both the nematic and the one-dimensional COs might be stabilized by tiny perturbations such as small magnetic fields or small strains \cite{Guo2024_NatPhys}. Nevertheless, whether these phases are stabilized by purely electronic interactions or by the strong electron-phonon coupling remain controversial \cite{Cho2020_NatComm,Tazai2022_arXiv,Grandi2024_arXiv}.

\noindent
By further lowering the temperature below T$_\text{sc} \sim 1$K, a superconducting state which might inherit the unconventional properties of the higher temperature phases is stabilized \cite{Ortiz2020_PRL,Ortiz2021_PRM,Yin2021_ChinPhysLett,Tazai2022_SciAdv,Guguchia2023_NatComm}. An inspection of the temperature vs pressure/doping phase diagram shows the presence of a double-dome superconductivity \cite{Chen2021_PRL,Wang2021_PRR,Du2021_PRB,Zheng2022_Nat,CapaSalinas2023_FEM,Yang2022_SciBull,Wu2024_ResSq}, with experimental evidence of different properties of this phase in the two domes \cite{Yang2022_SciBull}.

\section{Nonequilibrium control of kagome metals} \label{sec:noneq_control}
Various experimental threads have been pursued thus far to study kagome materials under non-equilibrium conditions, or to achieve time-dependent control specifically. Beyond the control aspect, nonequilibrium probes might be useful to gain information on some equilibrium properties of this class of compounds, such as their collective modes and the leading contribution to the mechanism that drives the CO.

\noindent
The properties of the CO stabilized below T$_\text{co}$ were analyzed by using coherent phonon spectroscopy \cite{Ratcliff2021_PRM}, finding a simultaneous condensation of three optical phonons in this phase, one related to an M-point instability and the other two coming from L-point instabilities. This result suggests a three dimensional character of the CO with $2 \times 2 \times 2$ reconstruction, in agreement with some theoretical analysis \cite{Christensen2021_PRB}. Interestingly, the experiment shows a transition to a metastable hidden state when the pump pulse hits the sample starting from a temperature close to T$_\text{co}$, related to a long ($\mu$s) recovery time of the original value of the reflectivity. This observation might be explained by a weak first-order character of the phase transition which suggests phase coexistence, in agreement with some experimental reports \cite{Mu2021_ChinPhysSoc,Li2022_NatComm,Song2022_SciChinPhys,Luo2022_npj}. Nevertheless, further analysis is required to better characterize the properties of the metastable state and to confirm this interpretation. By employing similar experimental techniques, the possibility to photoinduce a nonthermal modulation of the CO that can persist for nanoseconds was shown for CsV$_3$Sb$_5$ even far below the equilibrium critical temperature T$_\text{co}$ \cite{Yu2023_PRB}. By performing time-dependent density functional theory simulations, it was concluded that the nonthermal modulation of the CO comes from the optical tuning of the energy resonance between the band fermiology and the VHS present in the system \cite{Yu2023_PRB}.

\noindent
Additionally, the photoinduced melting and the subsequent recovery of the CO of CsV$_3$Sb$_5$ was recently studied by means of femtosecond time-resolved ARPES \cite{Azoury2023_PNAS}. This way, information regarding the amplitude component of the collective modes can be obtained, posing a lower boundary to the fastest time scale for the lattice reconstruction. From the analysis of the experimental data, it was concluded that a structural rather than an electronic mechanism is at the origin of the formation of the CO. With the same experimental technique, the dynamics of the band structure of CsV$_3$Sb$_5$ after a laser excitation was studied \cite{Zhong2024_arXiv}. They observe a sudden shift of the VHSs towards the Fermi level, followed by a decay that takes place in $\sim0.2$ps that happens together with an oscillation synchronous with a phonon mode at $1.3$THz. This lattice vibration, which is typically measured just in the CO phase, is observed even above T$_\text{co}$, suggesting the potential presence of CO fluctuations in this temperature regime, not in disagreement with other experimental findings \cite{Chen2022_PRL,Subires2023_NatComm,Yang2023_PRB}. The experimental report supports the presence of strong electron-phonon coupling in CsV$_3$Sb$_5$, as well as the possibility to tune the position of the VHS near the Fermi level at the ultrafast time scale.

\noindent
Another experiment has instead focused on the time-dependent changes in the superlattice peaks of the CO of CsV$_3$Sb$_5$ induced by a light pulse, monitored by time-resolved x-ray diffraction measurements. Interestingly, the coexistence of two distinct $2 \times 2 \times 2$ CO competing with each other is found. One of the two COs is more resistant than the other to photo-excitation, and it grows at the expenses of the other after the arrival of a pump. This experiment shows how nonequilibrium routes might be used to actually simplify the complex phase competitions among states which are strongly intertwined in equilibrium \cite{Basov2017_NatMat,delaTorre2021_RMP}.

\noindent
Recently, also the first out-of-equilibrium analysis of RbV$_3$Sb$_5$ has been conducted \cite{Xing2024_Nat}. By using a laser-coupled scanning tunneling microscopy,  a reversible switching of the relative intensity of the CO peaks, a property that is related to the chirality of the state, was shown employing linearly polarized light \cite{Jiang2021_NatMat,Shumiya2021_PRB,Wang2021_PRB}. A similar control over the intensity of the CO peaks was reported with magnetic fields, which requires the TRS breaking of the state. The chirality of the state, so far, has been experimentally detected only when the three independent $2 \times 2$ peaks have a different intensity. This can happen in two different cases, where one does not necessarily exclude the other: either because a nematic transition has occurred, leading to a breaking of the rotational part of the point group symmetry of the lattice, or because the one-dimensional $1 \times 4$ CO is stabilized. In the latter case, two new peaks corresponding to this additional translational symmetry breaking appear in diffraction measurements. Besides giving rise to these peaks, the onset of the $1 \times 4$ CO also affects the relative intensity of the $2 \times 2$ diffraction peaks. In particular, it is experimentally observed \cite{Xing2024_Nat} that the $2\times 2$ peaks that lay in the same reciprocal space direction as the $1 \times 4$ ones become larger than the others, which, instead, display a very similar, even if not equal, intensity. All in all, this study shows time-dependent control over the intensity of the subdominant $2 \times 2$ CO peaks by suitable linearly polarized light pulses. Interestingly, the same control over the relative intensities of the two CO peaks cannot be achieved with circularly polarized light.

\noindent
The works discussed above pave the way towards the time-dependent control of the properties of the vanadium-based kagome metals. For instance, it would be interesting to explore the time-dependent control of the TRS of the state in an experimental setup where nematicity or the $1 \times 4$ CO are absent, and to understand whether a circularly polarized pulse can induce chirality in the state, the control of which seems not allowed in the presence of the $1 \times 4$ CO. Also studying the time dependent manipulation of nematicity seems a tantalizing perspective and, to the best of our knowledge, would constitute the first realization of this kind of time-dependent control of this state of matter. In the next section, we provide a perspective on how these two time-dependent manipulation schemes might be achieved in the kagome metals.

\section{Time-dependent control of nematicity} \label{sec:td_nem}
On a lattice, nematicity constitutes a breaking of the rotational part of the point group symmetry of the lattice by the Fermi liquid formed by the electrons without any additional translation symmetry breaking, i.e., the nematic instability is a $\mathbf{q} = \mathbf{0}$ symmetry breaking. This symmetry breaking has been extensively studied for the iron-based superconductors, where some of the compounds belonging to this class show the onset of a nematic state in the proximity of a spin density wave (SDW) dome in the doping versus temperature phase diagram \cite{Fernandes2012_SuScTe}. The origin of this unconventional state might be related to the strong spin fluctuations expected in the proximity of the SDW \cite{Fernandes2012_PRB} which are argued to drive unconventional  superconductivity in these compounds \cite{Christianson2008_Nat,Chubukov2008_PRB}. The lattice of the iron-based superconductors is characterized by a C$_4$ rotational symmetry, and the nematic phase reduces it to C$_2$. Since there are two independent ways to move from C$_4$ to C$_2$, the nematic order parameter breaks an Ising $\mathbb{Z}_2$ symmetry.

\noindent
In the kagome metals, each kagome plane is characterized by a C$_6$ rotational symmetry, and the nematic phase which is found below T$_\text{nem}$ reduces it to C$_2$. As a consequence, the nematic order parameter belongs to the two-dimensional E$_2$ irreducible representation, i.e., the nematic order parameter has two components $\mathbf{N}^t = (N_1, N_2)$ differently from the case of a lattice with point group symmetry D$_{4 h}$. Furthermore, the nematic order breaks a Potts $\mathbb{Z}_3$ symmetry related to the three independent directions of the kagome lattice \cite{Grandi2024_arXiv}. The effective free energy for the nematic order of the kagome metals can be derived \cite{Grandi2023_PRB,Grandi2024_arXiv} or obtained from symmetry arguments, leading to:
\begin{align} \label{eq:free_en_nem}
	\mathcal{F}_\text{n} & = \frac{\alpha_\text{n}}{2} \mathbf{N}^2 - \frac{\gamma_\text{n}}{3} N_1 (N_1^2 - 3 N_2^2) + \frac{\beta_\text{n}}{4} \big( \mathbf{N}^2 \big)^2 ,
\end{align}
where $\alpha_\text{n}$ is the temperature-dependent quadratic coefficient, $\gamma_\text{n}$ the cubic and $\beta_\text{n}$ the quartic one. Assuming the system is not chiral and that it does not break the TRS, the potential in Eq.~\eqref{eq:free_en_nem} fully characterizes the properties of the nematic CO. One can imagine a full-optical control of nematicity by considering a two-pulse experimental setup, where the first unpolarized pulse drastically suppresses the nematic order while the second linearly polarized pulse guides the recovery of the order parameter in one of the three minima of the equilibrium free energy, similarly to what was proposed in the case of orbital ordering \cite{Hohenberg1977_RMP,Grandi2021_PRB} or for the control of chiral unconventional superconductivity \cite{Claassen2019_NatPhys,Yu2021_PRL}. Given the separation of the energy scales between the CO (T$_\text{co} \sim 90$K) and the nematic state (T$_\text{nem} \sim 30 - 50$K), it should be possible to suppress the latter without significantly altering the former.

\begin{figure}
    \centerline{\includegraphics[width=0.48\textwidth]{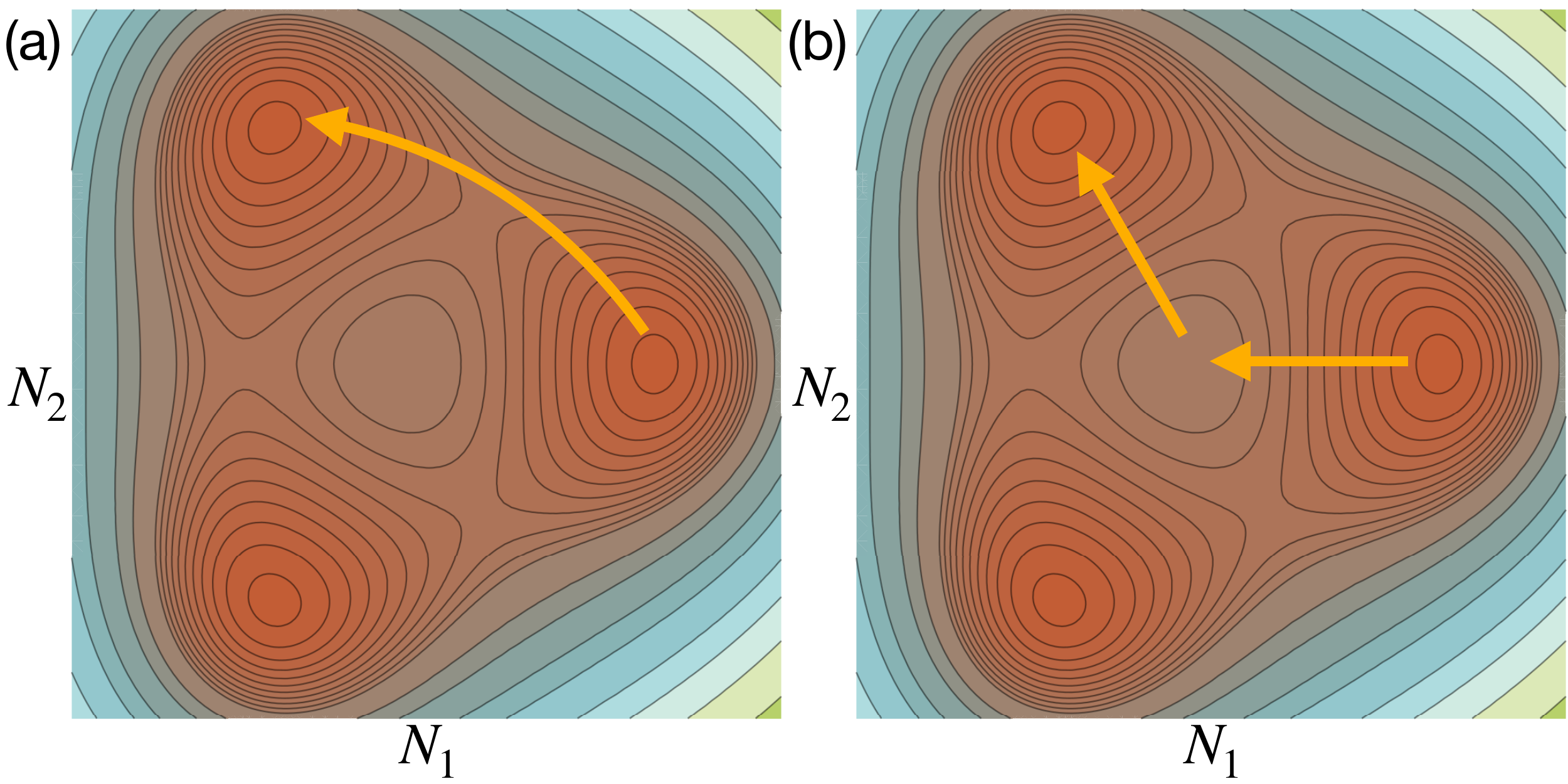}}
	\caption{ \textbf{Time-dependent control of nematicity -} Nematic free-energy landscape as a function of the two components of the nematic order parameter $N_1$ and $N_2$ ($\alpha_\text{n} = -1.3$, $\gamma_\text{n} = 0.5$ and $\beta_\text{n} = 0.8$). The color map goes from bright red to azure and finally to green by moving from low to high values of the potential. (a) Putative path for the switch of the nematic order by the action of a strain pulse. (b) Putative path for the switch of nematic order including the action of a light pulse that partially melts the order parameter combined with the action of the strain pulse that guides the recovery towards a different minimum.
    } \label{fig:Z3_pot}
\end{figure}

\noindent
When a coupling to a stress field $\boldsymbol{\sigma}_{\text{E}_2}^t = (\sigma_{\text{E}_{2,1}}, \sigma_{\text{E}_{2,2}})$ is added, the free energy obtained by integrating out the elastic deformations $\boldsymbol{\epsilon}_{\text{E}_2}^t = (\epsilon_{\text{E}_{2,1}}, \epsilon_{\text{E}_{2,2}})$ gets the additional contributions:
\begin{align} \label{eq:free_en_nem_strain}
	\tilde{\mathcal{F}}_\text{n} = \mathcal{F}_\text{n} & - \frac{\tilde{g}^2}{4 c_{\text{E}_2}} \mathbf{N}^2 - \frac{1}{4 c_{\text{E}_2}} \Big( g_1^2 \sigma_{\text{E}_{2,1}}^2 + g_2^2 \sigma_{\text{E}_{2,2}}^2 \Big) \nonumber \\
    &- \frac{\tilde{g}}{2 c_{\text{E}_2}} \Big( g_1 \sigma_{\text{E}_{2,1}} N_1 + g_2 \sigma_{\text{E}_{2,2}} N_2 \Big) ,
\end{align}
where $\tilde{g}$ is the magnitude of the nematoelastic coupling, $g_1$ ($g_2$) is the coupling constant between $\sigma_{\text{E}_{2,1}}$ ($\sigma_{\text{E}_{2,2}}$) and $\epsilon_{\text{E}_{2,1}}$ ($\epsilon_{\text{E}_{2,2}}$) and $c_{\text{E}_2}$ is the $\text{E}_2$ element of the stiffness matrix. The direct coupling between the strain and the nematic order parameter in Eq.~\eqref{eq:free_en_nem_strain} suggests that one might control or even induce nematicity with a suitable action of the strain fields, extending the applicability of strain engineering. This is not surprising: The strain induces a change in the lattice through the elastic deformations which, in turn, are also coupled to the electronic nematic order parameter. The nonequilibrium generalization of this way of controlling collective excitations and orders in materials is in its infancy \cite{Afanasiev2014_PRL,Baldini2019_SciAdv,Demenev2019_PRB,Kuznetsov2019_PRR}. To provide an example of the usage of this kind of control technique, the combination of optical and strain pulses has been used to control the insulator-to-metal transition in vanadium dioxide \cite{Mogunov2019_PRApp}.

\begin{figure*}
    \centerline{\includegraphics[width=0.99\textwidth]{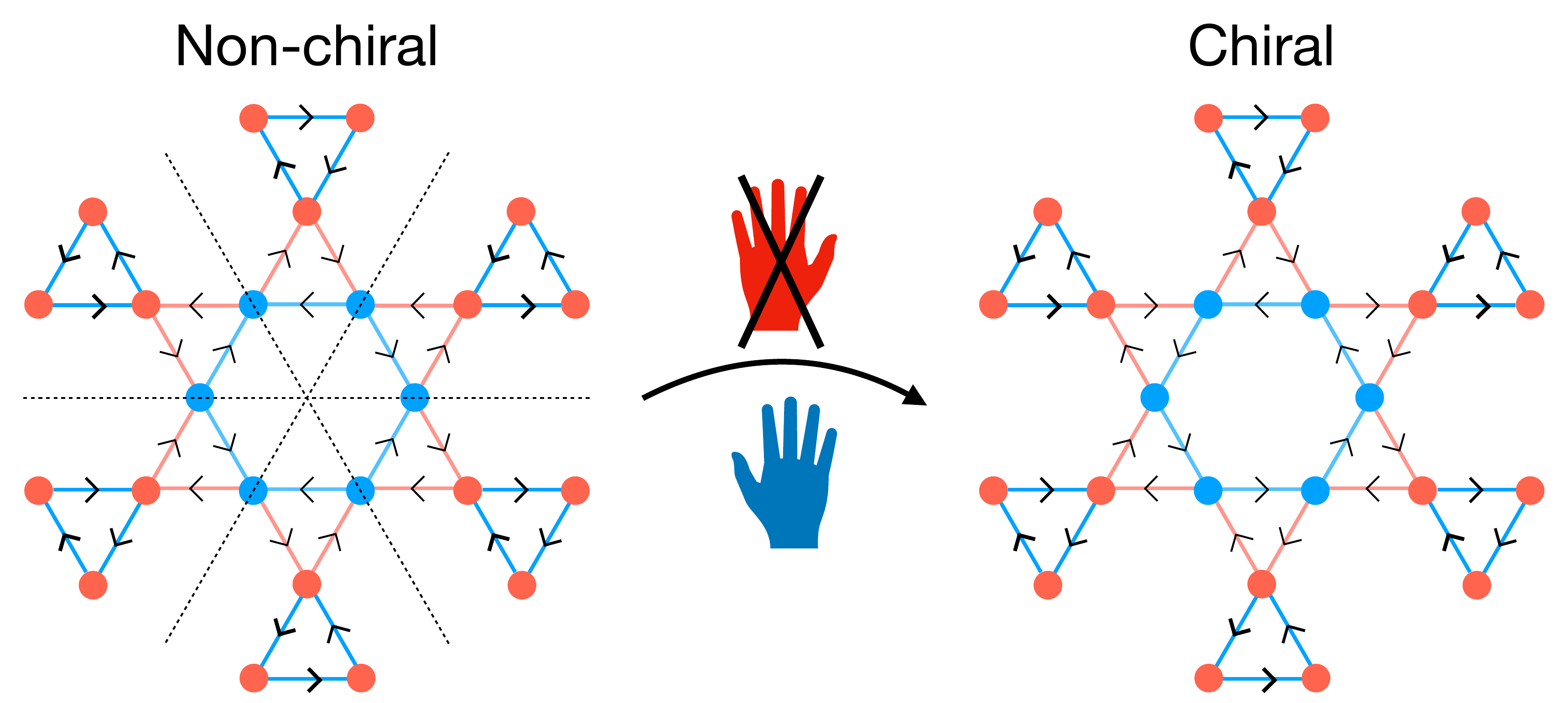}}
	\caption{ \textbf{Chirality induced by circularly polarized light -} Scheme for the putative light-induction of chirality in the $2 \times 2$ CO state of the kagome metals by a circularly polarized pulse with given handedness. Orange and azure sites (bonds) represent higher and lower occupations (bond strength), respectively. The arrows on the bonds show the direction of the current on that bond. On the left, the three dashed lines show the mirror planes of the initial non-chiral configuration. On the right, a chiral state is induced by the light pulse of given handedness (blue hand). By changing the handedness of the circular pulse (red hand), a chiral state with opposite chirality is reached.
    } \label{fig:chirality}
\end{figure*}

\noindent
The realization of such a strain pulse might take place in several ways \cite{Ruello2015_Ultrason}, and this should allow a time-dependent coupling to the nematic order parameter in kagome metals. Depending on the relative amplitude between the pulse intensity and the energy barrier the system has to overcome to switch minima in the nematic free energy landscape Eq.~\eqref{eq:free_en_nem_strain}, the strain pulse might be enough to induce this transition in the system, see Fig.~\ref{fig:Z3_pot}(a), or not. In the latter case, a more suitable control choice might be to first apply an optical pump with the aim of transiently heating the material, i.e., to partially melt the nematic order, and subsequently then to apply the strain pulse. In analogy with what has been predicted for orbital ordered systems \cite{Grandi2021_PRB}, the switching from one minimum to the other might become easier in this second case because of the transient excitation of the fluctuations of the order parameter. Indeed, since fluctuations already play an important role in equilibrium for the kagome metals, they might be even more relevant out of equilibrium, leading to a transient renormalization of the energy landscape that makes the switching easier, see the scheme shown in Fig.~\ref{fig:Z3_pot}(b) \cite{Dolgirev2020_PRB,Grandi2021_PRB}. This way, one might achieve a time-dependent control of nematicity in the kagome metals. We further stress the relevance of the two components of the nematic order parameter for the realization of the two time-dependent protocols sketched in Fig.~\ref{fig:Z3_pot}, which would not be possible to realize with an Ising-like single component order parameter like in pnictides.

\section{Time-dependent control of chirality and of time reversal symmetry} \label{sec:td_chi}
Chirality is a property displayed by systems that cannot be mapped to themselves by a combination of rotations and translations after the action of a mirror symmetry. Several experimental studies found a significant chiral transport in structurally chiral materials, which are, nevertheless, difficult to be externally controlled and switched. Recent advancements have shown that it is possible to photo-induce a chiral state in the structurally achiral (although antiferro-chiral) compound boron phosphate by the excitation of specific phonon displacements that increase and reduce the magnitude of the two chiral domains \cite{Zeng2024_arXiv}. In the kagome metal CsV$_3$Sb$_5$, which is structurally achiral and centrosymmetric, chiral transport has been measured via second-harmonic generation with in-plane magnetic fields below $\sim 35$K. Particularly, a change of the chirality from left to right handed can be induced by changing the sign of the out-of-plane magnetic field \cite{Guo2022_Nat,Guo2024_npjQM}. These two recent advancements put the kagome metals at the forefront for the realization of chiralitronic devices \cite{Taylor2019_APL,Taylor2020_PRB}, which puts forth the vision of being able to control this property on the ultrafast time-scale.

\noindent
The TRS is instead the discrete symmetry that maps one state to its partner in which, however, the time is flowing backwards. In the case of a state with orbital currents, it would correspond to the configuration with all the currents flowing in the opposite direction. Generally speaking, TRS and chirality are independent symmetry breaking, and both are characterized by a discrete $\mathbb{Z}_2$ Ising variable.

\noindent
Several experiments have found a broken TRS at the onset of the $2 \times 2$ CO. The relation between the broken TRS and the $2 \times 2$ CO is rooted in the nature of the CO of the vanadium-based kagome metals, which is likely a charge bond order (CBO) with three independent intersite order parameters $\Delta_1$, $\Delta_2$ and $\Delta_3$ \cite{Kiesel2013_PRL,Wang2013_PRB,Profe2024_arXiv} that start having an imaginary component that breaks the TRS as soon as they become finite \cite{Feng2021_PRB}. However, imaginary values of the three order parameters do not necessarily lead to a chiral state \cite{Denner2021_PRL,Grandi2023_PRB}. The observation of a chiral response only at low temperatures (below $\sim 35$K) might suggest a correlation between this property and the onset of the one dimensional $1 \times 4$ CO \cite{Guo2022_Nat} or of nematicity \cite{Jiang2021_NatMat}. Since the presence of the $1 \times 4$ CO also leads to asymmetric $2 \times 2$ peaks as for the nematic state, chirality sets in, generally speaking, when the three order parameters that characterize the $2 \times 2$ CO start having a different amplitude or a different phase. Indeed, in this case, it is possible to properly define a clock- or anticlockwise counting direction going from the lowest to the highest peaks of the Fourier transform in scanning tunneling microscopy maps \cite{Jiang2021_NatMat}.

\noindent
In equilibrium the TRS breaking and the chirality of the state are disentangled one from the other, i.e., in general, one does not imply the other. However, a suitable perturbation of the system might entangle the two symmetry breakings. This is the case for a circularly polarized light pulse, which simultaneously breaks the TRS and it has a defined handedness. The action of a pulse of this kind starting from the $2 \times 2$ CO above $\sim 35$K, i.e., in the temperature-window in which the state breaks only the translational symmetry of the pristine lattice and the TRS, might lead to a change over the distribution of the imaginary part of the CBO \cite{Grandi2023_PRB}, putatively leading to the transient stabilization of a metastable state. Moreover, the chirality of the pulse might induce the same property in the material over which the pulse is shone on, see Fig.~\ref{fig:chirality}. This way, it might be possible to disentangle by means of nonequilibrium driving the intertwining between chirality and nematicity/$1 \times 4$ CO observed in equilibrium. In a similar spirit of  Floquet engineering already employed for graphene \cite{McIver2020_NatPhys}, one might even try to induce unconventional (topological) states different from the equilibrium counterpart by shining a circularly polarized pulse over the kagome metals above T$_\text{co}$, allowing us to verify the theoretical predictions that have been made so far for the Floquet driving of the kagome lattice \cite{He2014_PhysLettA,Du2018_PRB,Ohgushi2000_PRB,Liu2022_PRB}.

\section{Conclusions} \label{sec:concl}
As outlined in the previous sections of this perspective article, the kagome metals, and especially the vanadium-based 135 family, have established themselves as a very promising platform to study the competition between several unconventional states such as a CO with broken time-reversal symmetry without local magnetic moments, nematicity, a chiral state putatively related to nematicity or to the onset of a one-dimensional CO and unconventional superconductivity. First, additional experimental studies are required to clarify the equilibrium phase diagram of these materials, which is still intensely discussed. This task might be complicated by the likely high-sensitivity of the kagome metals, and especially of CsV$_3$Sb$_5$, to external perturbations \cite{Guo2024_NatPhys}. This feature suggests a rather rich energy landscape for these systems, which makes them perfect experimental platforms for studying the nonequilibrium control of their properties, as hinted at by recent experiments \cite{Xing2024_Nat}.

\noindent
We have outlined the possibility to control the nematic order parameter in a Ginzburg-Landau framework thanks to the combined action of light and strain pulses. Moreover, we have stressed the relevance of Floquet driving either to disentangle chirality and nematicity/the one-dimensional charge order out-of-equilibrium or to induce a nontrivial topological state different from its equilibrium counterpart. Some of these ideas might be applied, \textit{mutatis mutandis}, also to other classes of kagome metals such as the 135 titanium-based family, which does not show any CO but only nematicity and a putative unconventional superconducting state \cite{Yang2022_arXiv,Hu2023_NatPhys}.

\begin{acknowledgements}
The authors declare no conflicts of interest. F.G. and D.M.K. acknowledge support by the DFG via Germany’s Excellence Strategy$-$Cluster of Excellence Matter and Light for Quantum Computing (ML$4$Q, Project No. EXC $2004/1$, Grant No. $390534769$), individual grant No. 508440990 and within the Priority Program SPP 2244 ``2DMP'' --- 443274199. This work was supported by the Max Planck-New York City Center for Nonequilibrium Quantum Phenomena. R.T. received funding from the Deutsche Forschungsgemeinschaft (DFG, German Research Foundation) through Project-ID $258499086$-SFB $1170$ and through the W\"{u}rzburg-Dresden Cluster of Excellence on Complexity and Topology in Quantum Matter - \textit{ct.qmat} Project-ID $390858490$ - EXC $2147$.
\end{acknowledgements}


\end{document}